\documentclass[a4paper,amsmath,amssymb]{jpconf}
\usepackage{graphicx}
\usepackage{amsmath, amsthm, amssymb, color, subfigure}
\begin{document}
\title{Spatiotemporal pattern formation in a prey-predator model under environmental driving forces}

\author{Anuj Kumar Sirohi}

\address{School of Computational and Integrative Sciences, Jawaharlal Nehru University, New Delhi-110067, INDIA
}
\ead{anujsirohi.amu@gmail.com}

\author{Malay Banerjee}

\address{Department of Mathematics and Statistics,
Indian Institute of Technology,
Kanpur-208016, INDIA
}

\ead{malayb@iitk.ac.in}

\author{Anirban Chakraborti}

\address{School of Computational and Integrative Sciences, Jawaharlal Nehru University, New Delhi-110067, INDIA
}
\ead{anirban@jnu.ac.in}

%%%%%%%%%%%%%%%%%%%%%%%%%%%%%%%%%%%%%%%%%%%%%%%%%%%%%%%
%%                                                                               Abstract                                                                                             %%
%%%%%%%%%%%%%%%%%%%%%%%%%%%%%%%%%%%%%%%%%%%%%%%%%%%%%%%
\begin{abstract}
Many existing studies on pattern formation in the reaction-diffusion
systems rely on deterministic models. However, environmental noise
is often a major factor which leads to significant changes in the
spatiotemporal dynamics. In this paper, we focus on the
spatiotemporal patterns produced by the predator-prey model with
ratio-dependent functional response and density dependent death rate
of predator. We get the reaction-diffusion equations incorporating
the self-diffusion terms, corresponding to random movement of the
individuals within two dimensional habitats, into the growth
equations for the prey and predator population. In order to have to have the 
noise added model, small amplitude
heterogeneous perturbations to the linear intrinsic growth rates are
introduced using uncorrelated Gaussian white noise
terms. For the noise added system, we then observe
spatial patterns for the parameter values lying
outside the Turing instability region. With thorough numerical
simulations we characterize the patterns corresponding to Turing and
Turing-Hopf domain and study their dependence on different system
parameters like noise-intensity, etc.
\end{abstract}
%%%%%%%%%%%%%%%%%%%%%%%%%%%%%%%%%%%%%%%%%%%%%%%%%%%%%%%%%
\section{Introduction}
\label{sec:Introduction}
\par
\medskip

Self-organizing spatial pattern formation in interacting population
models is an important area of research to understand the
distribution of the species within their habitats. Mainly
reaction-diffusion equation models are studied analytically with
supportive numerical simulations to understand the nature of
population patches and whether they are changing with respect to
time or not. Investigations for various types of Turing and
non-Turing pattern formations in two dimensional prey-predator
models have received significant attention \cite{alonso2002mutual, banerjee2011self,
baurmann2007instabilities, medvinsky2002spatiotemporal, Wang2007spatio}. Spatio-temporal prey-predator models exhibit various types of self-organizing spatial patterns; either
they are stationary or non-stationary. Stationary patterns are
characterized by heterogeneous distribution of populations over
space which do not change with time once they reach the non-constant
steady-state. On the other hand, the non-stationary patterns
change with time continuously and never reach any non-constant
steady-state. The stationary patterns produced by the
spatio-temporal prey-predator models for two dimensional spatial
domains can be classified as: (i) spot patterns
(cold-spots and hot-spots), (ii) labyrinthine/stripe patterns, or (iii) the mixture of
spot and stripe patterns. These three types of patterns are mostly
observed for parameter values taken from Turing-domain or
Turing-Hopf-domain, when  the  diffusivity of two species differ by a
significant magnitude. Sometimes one does find labyrinthine patterns
as a non-stationary pattern but for parameter values well inside
the Turing-Hopf domain. The non-stationary patterns may include periodic
patterns (sometimes as periodic traveling waves), interacting
spiral patterns and chaotic patterns. The occurrences of
spatio-temporal chaos for parameter values lying within the Turing
domain or outside the Turing-domain remains a debatable and controversial issue
\cite{malchow2007spatiotemporal}. There is no unified approach in determining the
parameter values in a systematic way or in identifying the mechanism which
leads to the spatio-temporal chaotic patterns \cite{banerjee2011self,
banerjee2012turing}. Derivation of necessary-sufficient condition(s),
either in terms of parametric restrictions or system
characterization for the onset of spatio-temporal chaotic patterns
for prey-predator type interacting models  thus remains an open
problem.

Recently, the role of environmental noise on spatio-temporal pattern
formation has been studied by the researchers for the noise added
spatio-temporal models of interacting populations
\cite{malchow2004oscillations, sun2009role, Wang2011color, li2012pattern}. In most of the cases, the
authors have been used temporally correlated and spatially
uncorrelated coloured noise terms. Although in some investigations
the alteration of spatial patterns due to the presence of additive
coloured noise terms have been reported but this type of formulation
is not ecologically viable as they correspond to spontaneous
generation or inhibition of populations at the places with no
population or very low concentration of population. On contrary, the
consideration of multiplicative noise terms signify the noisy
variation in some system parameters involved with the reaction
kinetics. It is evident that due to the environmental fluctuations,
the carrying capacity, birth and death rates, intensity of intra-
and inter-species competition rates, predation rates exhibit random
fluctuations to a modular extent around some average value
\cite{may1972limit}. Perturbation of one or more rate constants by the noise
terms leads to spatio-temporal models with multiplicative noise
terms. Several noise induced phenomena like noise-induced phase
transitions, noise enhanced stability/instability, noise delayed
extinction, stochastic resonance are investigated for the
spatio-temporal ecological models considered within fluctuating
environments. These analytical/numerical findings provide important
information towards how the species establish within their respective habitats
or help us to identify the important risk factors leading towards the
extinction of species.

The main objective of this paper is to study a simple model of 
predator-prey with ratio-dependent functional response and density 
dependent death rate of predator, along with the self-diffusion terms corresponding to the random movement of the individuals within two dimensional habitats, and study the influence of small amplitude
heterogeneous perturbations to the linear intrinsic growth rates.

%%%%%%%%%%%%%%%%%%%%%%%%%%%%%%%%%%%%%%%%%%%%%%%%%%%%%%%%%
\section{The spatiotemporal model}
%%%%%%%%%%%%%%%%%%%%%
\label{sec:Basic Model}

\medskip

\setcounter{equation}{0} % sets equation counter to 1

Consider that $u \equiv u(x,y,t)$ and $v \equiv v(x,y,t)$ represent the prey and predator population densities at any time $ t $ and at  the spatial location $ (x,y)\in \Omega $, where $ \Omega \in R^{2}$ is a square domain in $\mathbb{R}$ with boundary $\partial{ \Omega} $. 
The spatiotemporal dynamics of prey-predator interaction with logistic growth for prey, ratio-dependent functional response and density dependent death rate for predators is governed by the following system of nonlinear coupled partial differential equations (PDE), in terms of dimensionless variables and parameters:

\begin{eqnarray}
\frac{\partial u}{\partial t} & = & u(1-u) -\frac{\alpha uv}{u+v} + \nabla^{2} u \equiv f(u,v)+\nabla^{2}u,         \label{eq1}
\\
\frac{\partial v}{\partial t} & = & \frac{\beta uv}{u+v} - \gamma v - \delta v^{2} +d\nabla^{2} v \equiv g(u,v)+\nabla^{2}v
\label{eq2}
\end{eqnarray}
subjected to positive initial conditions:
\begin{eqnarray}
 u(x,y,0)\equiv u_0(x,y)>0, v(x,y,0)\equiv v_0(x,y)>0,\text{for all} (x,y)\in\Omega
\label{eq:equation3}
\end{eqnarray}
and zero-flux boundary conditions:
\begin{eqnarray}
 \frac{\partial u}{\partial \nu} = \frac{\partial v}{\partial \nu} = 0, \text{ for all} (x,y)\in \partial{ \Omega}, t>0
\label{eq:equation4}
\end{eqnarray}
where $\nu$ is the outward drawn unit normal vector on the boundary, and $ \alpha$, $\beta$, $\gamma$ and $\delta$ are positive dimensionless parameters. The ratio of diffusivity of two species is denoted as $ d $. 
The diffusion terms in this reaction-diffusion system for the prey and predator population are introduced to take care of the random movements of the individuals within two dimensional habitats. The zero-flux boundary condition indicates that predator-prey system is self-contained within the two dimensional habitat with no population flux across the boundary.
%%%%%%%%%%%%%%%%%%%%%%%%%%%%%%%%%  Sub Section %%%%%%%%%%%%%%%%%%%%%%%%%%%%%%%
\subsection{Turing Bifurcation}
\par
\medskip

The ecologically feasible and coexisting equlibrium point for the correspoding temporal model are spatially homogeneous steady state for the reaction diffusion model (\ref{eq1}-\ref{eq2}) and can be obtained by solving the following two algebric equations, with $u,v>0$:
\begin{eqnarray}
    f(u,v) & = & u(1-u) -\frac{\alpha uv}{u+v}  = 0, \label{eq3} \\
    g(u,v) & = & \frac{\beta uv}{u+v} - \gamma v - \delta v^{2}  =  0.
\label{eq4}
\end{eqnarray}
If we denote the coexisting equilibrium point by $E_*\left(u_{*},v_{*}\right)$, then $u_*$ and $v_*$ satisfy the equtions $\displaystyle 1-u -\frac{\alpha v}{u+v}\,=\,0\,=\,\frac{\beta u}{u+v}-\gamma -\delta v$. The explicit expression for the components of $E_*$, conditions for their feasible existence and possible number of coexisting steady-states are discussed in Ref.~\cite{banerjee2012turing} extensively. Further we assume that $E_*$ is coexisting  homogeneous steady-state which is locally asymptotically stable for the temporal counterpart of the system
(\ref{eq1}-\ref{eq2}), i.e., the conditions $ a_{11} + a_{22}  <  0 $ and $ a_{11} a_{22} - a_{12} a_{21} >  0 $ are simultaneously satisfied, where $a_{11}, a_{12}, a_{21} $ and $ a_{22}$ are given by
\begin{eqnarray*}
    a_{11} \,=\,  1 -2u_* -\frac{\alpha v_*^{2}}{(u_*+v_*)^{2}}, \,\,\,
    a_{12} \,=\,  - \frac{\alpha u_*^{2}}{(u_*+v_*)^{2}}, \,\,\,
    a_{21}  \,=\, \frac{\beta v_*^{2}}{(u_*+v_*)^{2}}, \\
    a_{22} \,=\,  -\gamma-2\delta v_* + \frac{\beta u_*^{2}}{(u_*+v_*)^{2}}. \hspace{1.5in}
\end{eqnarray*}
\par

The basic criteria for the diffusion-driven instability or Turing instability is that the stable homogeneous steady-state becomes unstable due to small amplitude heterogeneous perturbation around the homogeneous steady-state. Hence for the occurrence of Turing instability, first we need a locally stable coexisting steady-state for the corresponding temporal model. The linear stability analysis for spatiotemporal model (\ref{eq1}-\ref{eq2}) around the homogeneous steady-state $E_*$ and the conditions required for the onset of Turing-instability are well established and used by several researchers. Here, we just mention the three basic mathematical criteria required for Turing-instability of the model (\ref{eq1}-\ref{eq2}) around $E_* (u_{*}, v_{*} )$, as follows:\\
\begin{eqnarray}
    a_{11} + a_{22} & < & 0, \label{Tur1}\\
    a_{11} a_{22} - a_{12} a_{21}  & > & 0,  \label{Tur2}        \\
    da_{11} + a_{22}  & > &2\sqrt{d}\sqrt{a_{11} a_{22}- a_{12} a_{21} }, \label{Tur3}
\end{eqnarray}
and Turing-instability sets in at the critical wave number
\begin{eqnarray}
    k_{cr}^2 = \frac{d a_{11} + a_{22}}{2\sqrt{d} } > 0.
\label{Turkcritical}
\end{eqnarray}
Satisfaction of the condition (\ref{Tur3}) implies homogeneous steady-state becomes unstable under small amplitude heterogeneous perturbation for wavenumbers $k>k_{cr}$ and Turing pattern emerges. Type(s) of emerging spatial pattern(s) (as mentioned earlier in the introduction), depend upon the specific choice of parametric values and magnitude of $ d $.
\par The Turing-bifurcation curve is defined by the following equation
$$ da_{11} + a_{22}  \,=\, 2\sqrt{d}\sqrt{a_{11} a_{22}- a_{12} a_{21} }.$$
\par Solving this equation for $  d $, we can find the critical threshold for the ratio of diffusivity $d_{cr}$, above which the Turing patterns emerge. In the next section we are going to report the varieties of spatial patterns generated by the system (\ref{eq1}-\ref{eq2}) for suitable choice of
the parameter values satisfying the Turing bifurcation conditions.
%%%%%%%%%%%%%%%%%%%%%%%%%%%%%%%%%%%%%%%%%%%
\section{Self-organizing spatial patterns}
\medskip

In this section, we present the numerical simulation results and varieties of spatial patterns  that are exhibited by the spatiotemporal model (\ref{eq1}-\ref{eq2}) for parameter values
within the Turing and Turing-Hopf domain. If we take the parameter
values $\alpha = 2.0$, $\beta = 1$, $\gamma = 0.6$ and $\delta =
0.1$ then we find the unique coexisting equilibrium point
$E_*(0.2287,0.1436)$ which is a homogeneous steady-state for the
system (\ref{eq1}-\ref{eq2}). $E_*$ is locally asymptotically stable
for the temporal counterpart as the conditions (\ref{Tur1}) and
(\ref{Tur2}) are satisfied. In order to understand the role of
diffusivity of the prey and the predator towards the destabilization
of the homogeneous steady-state we need to find the quantity
$d_{cr}$ as mentioned in the previous section. For the chosen set of parameter
values, we find $d_{cr}=4.9157$ and hence the heterogeneous
perturbations lead to Turing patterns for $d>d_{cr}$.

\par
\begin{figure}%
\leavevmode \centering
\includegraphics[]{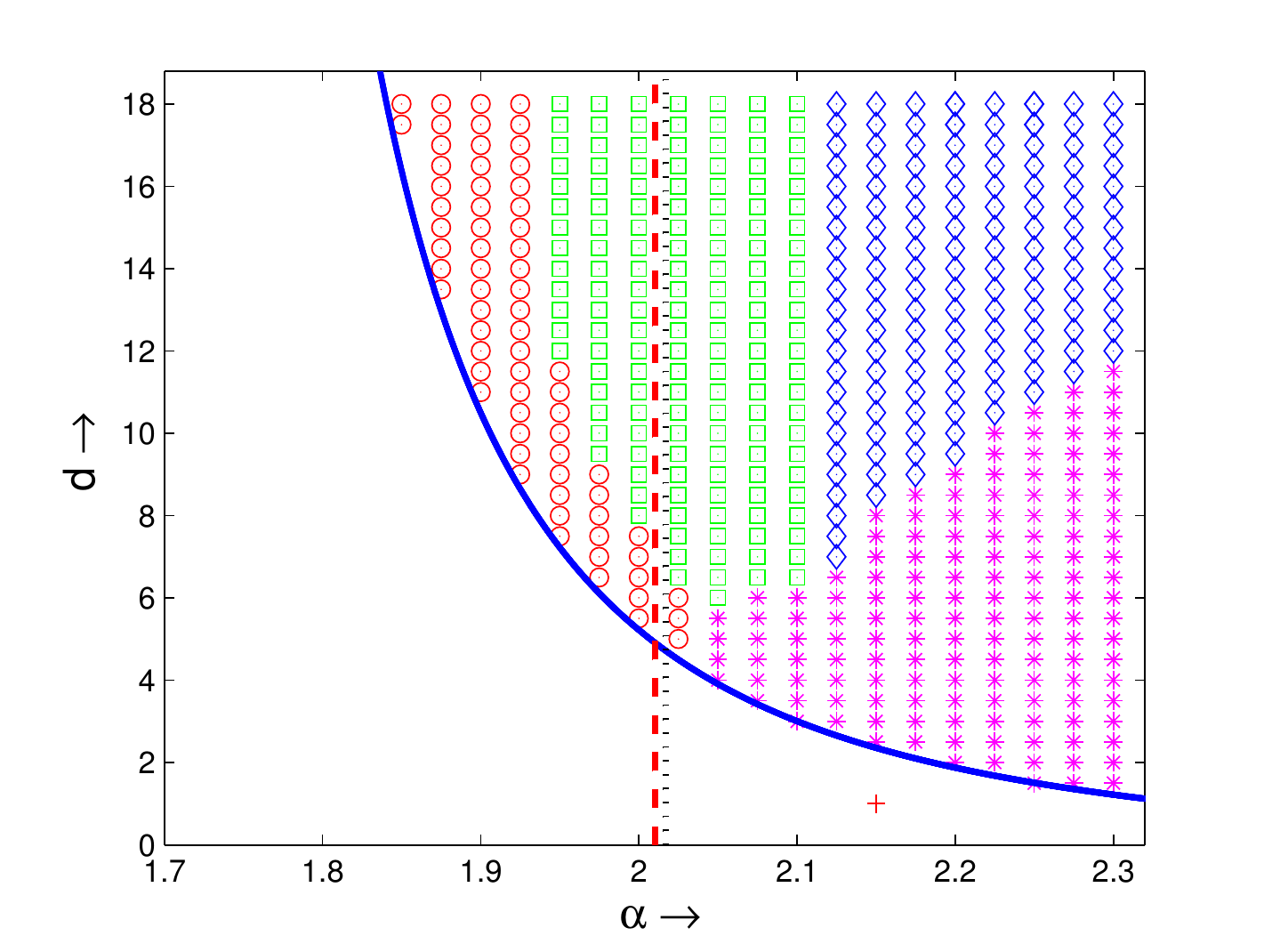}
\caption{\footnotesize Five different types of spatial patterns
observed for the parameter values lying within the Turing
instability region (above the blue curve which is Turing bifurcation
curve) and below Turing instability region. Choices of parameter values are marked with four different symbols based upon the resulting patterns:
\textcolor{red}{${\circ}$} cold-spot, \textcolor{green}{$\Box $}
mixture of spot-stripe, \textcolor{blue}{$\diamond$} labyrinthine, \textcolor{magenta}{$\ast$} chaotic and \textcolor{red}{${+}$} interacting spiral. Hopf-bifurcation curve (shown by red dashed line) and temporal homoclinic
bifurcation curve (shown by black dotted curve).}\label{fig:1}
\end{figure}

 The existence and non-existence of Turing patterns are solely
dependent upon the magnitude of the parameters involved with the
model under consideration. Here we consider $\alpha$ and $d$ as the
bifurcation parameters to construct the Turing bifurcation diagram.
We have presented the Turing bifurcation diagram in $\alpha
d$-parametric plane (see~Fig.~\ref{fig:1}). The bifurcation
diagram is prepared keeping $\beta = 1$, $\gamma = 0.6 $ and $\delta
= 0.1$ fixed. $\alpha$ and $d$ are considered as the controlling
parameters to obtain different spatiotemporal patterns. We have
presented three bifurcation curves in the bifurcation diagram,
namely Turing-bifurcation curve (blue curve), temporal
Hopf-bifurcation curve (red dashed line) and temporal homoclinic
bifurcation curve (black dotted curve). The coexisting equilibrium point
$E_{*}$ for the temporal model, corresponding to the system
(\ref{eq1}-\ref{eq2}), is stable for $\alpha<\alpha_{h}$ and it
looses stability through the Hopf-bifurcation at $\alpha_{h} = 2.01$
(approx). The Turing instability region is the region lying above
the Turing bifurcation curve and which is divided into two parts by
the vertical line $\alpha=\alpha_{h}$. Part of the Turing domain
lying in the region $\alpha>\alpha_{h}$ is the Turing-Hopf domain
where temporal and spatiotemporal perturbations are both unstable.

\par For the chosen set of parameter values we find only one feasible
coexisting equilibrium point $E_{*}$ therefore we will discuss only
the spatial-pattern formation around this homogeneous steady-state.
We have obtained various patterns generated by the model under
consideration through exhaustive numerical simulations. All
numerical simulations are carried out over a $200\times200$ lattice
with time-step $\Delta t = 0.01$ and spatial steps $\Delta x =
\Delta y = 2$. All numerical simulations are performed using the
forward Euler method for the temporal part and five point finite
difference scheme for the diffusion part. A small amplitude
heterogeneous perturbations to the homogeneous steady-state are
introduced by $ u_{0}\left( x_{i},y_{j}\right) = u_* + \epsilon
\xi_{ij}$, $ v_{0}\left( x_{i},y_{j}\right) = v_* + \epsilon
\eta_{ij}$ where $ \epsilon = 0.001 $, $\xi_{ij}$ and  $\eta_{ij}$
$(i, j = 1, 2)$ are spatially uncorrelated Gaussian white noise
terms. We have observed four different types of patterns within the
Turing domain, cold-spot, labyrinthine, mixture of spot and stripe
and spatiotemporal chaotic patterns.

%%%%%%%%%%%%%%%%%%%%%%%%%%%%%%%%%%%%%%%%%%%%%%%%%%%%%%%%%%%%%%%%%
%%%%%%%%%%%%%%%%%%%%%%%%%%%%%%%%%%%%%%%%%%%%%%%%%%%%%%%%%%%%%%%%%%%%

\par We have observed only cold-spot and mixture of spot-stripe patterns
for parameter values within the Turing domain, whereas mixture of
spot-stripe, labyrinthine and chaotic patterns are observed for
parameters lying in the Turing-Hopf domain. Numerical simulations have been  performed for systematic choices of $\alpha$ and $d$ within the Turing domain in order to understand the entire variety of spatiotemporal patterns for parameter values within the Turing domain. Obtained patterns for different combinations of $\alpha$ and
$d$ are marked with four different symbols. The whole regime of
patterns in the Turing domain is shown in the Fig.~\ref{fig:1}.

%%%%%%%%%%%%%%%%%%%%%%%%%%%%%%%%%%%%%%%%%%%%%%%%%%%%%%%%%%%%%%%%
\begin{figure}%
\leavevmode \centering
\mbox{\subfigure[]{\includegraphics[width=2.5in,height=2in]{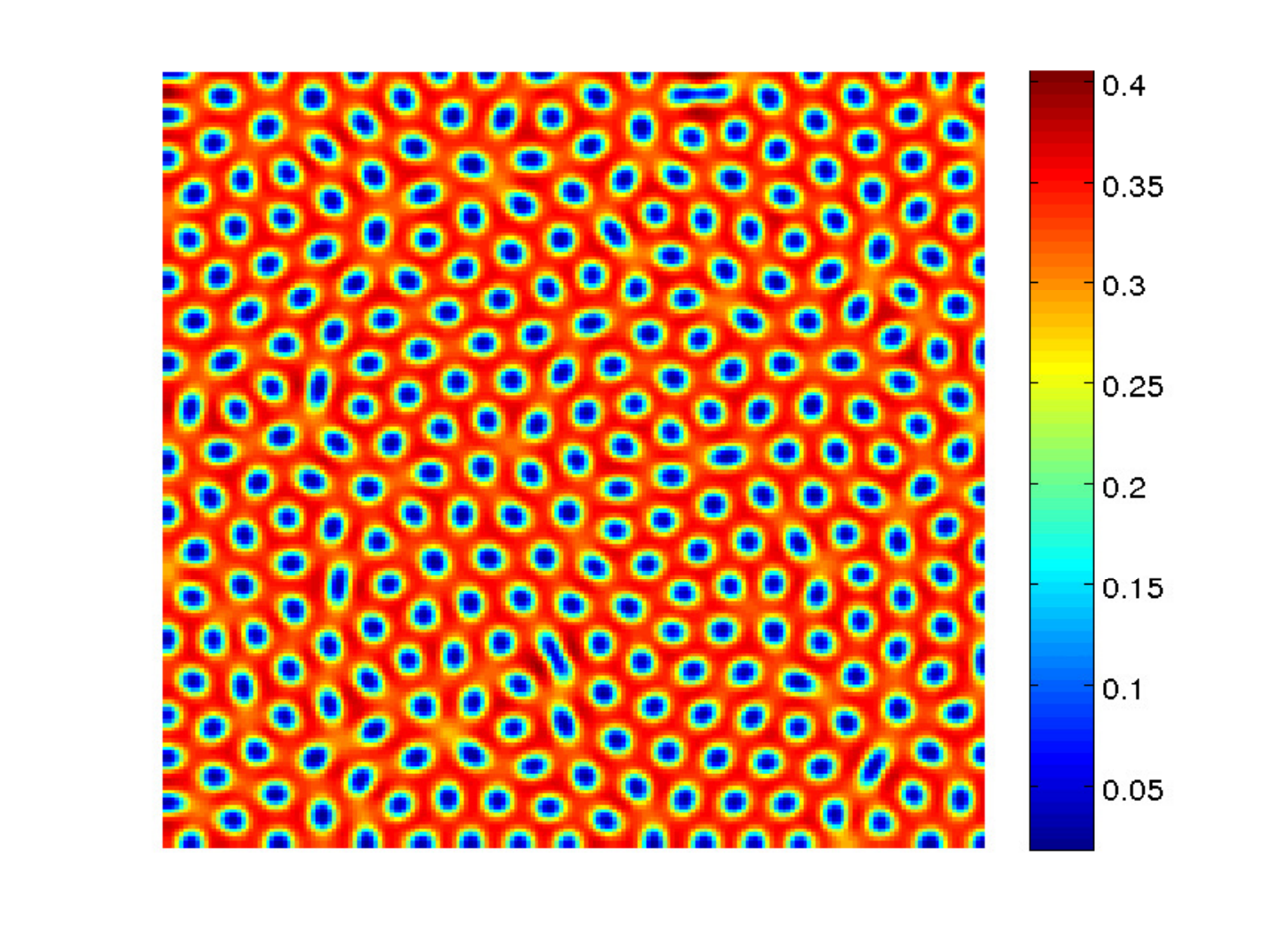}}
\subfigure[]{\includegraphics[width=2.5in,height=2in]{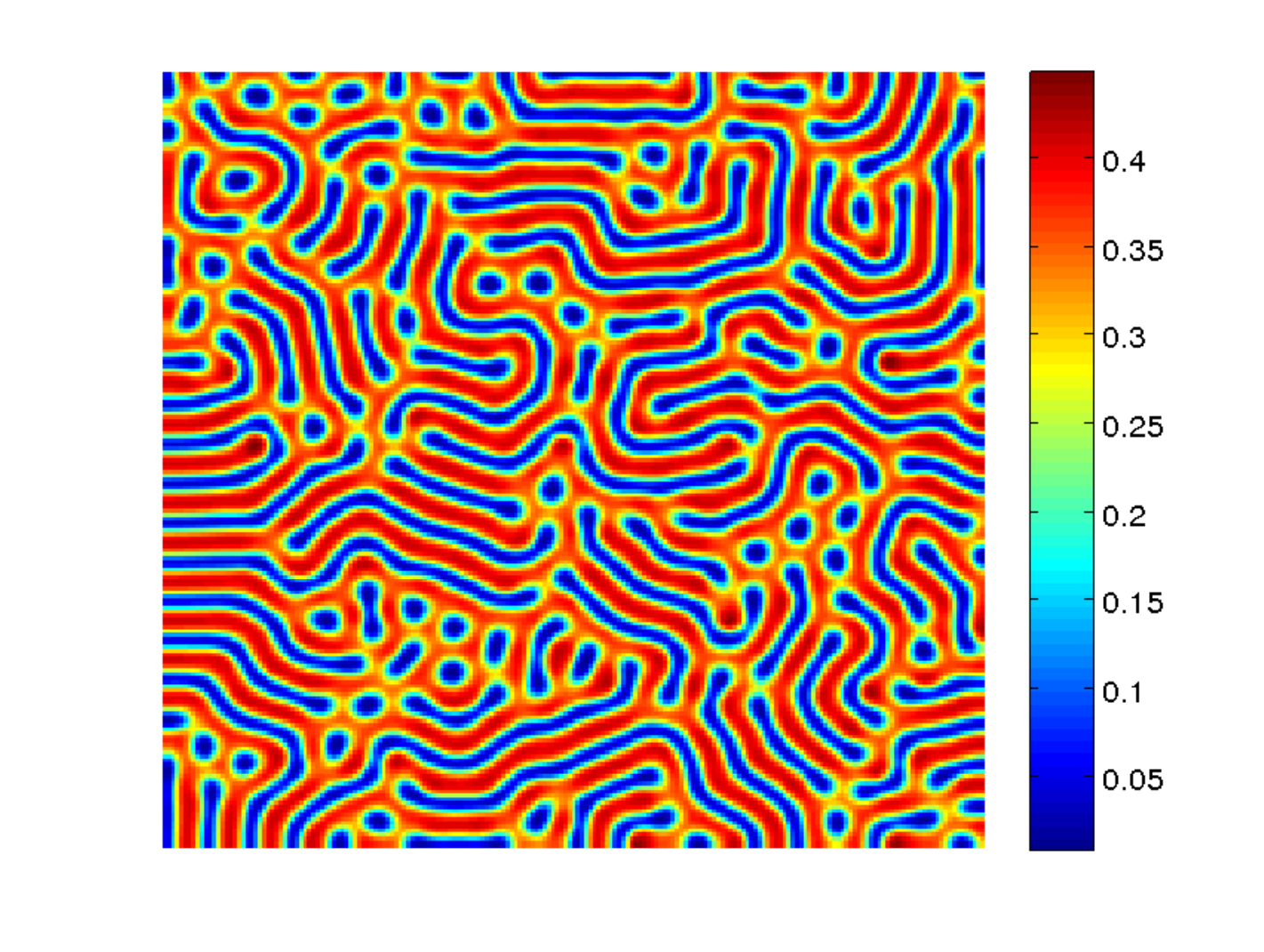}}}
\mbox{\subfigure[]{\includegraphics[width=2.5in,height=2in]{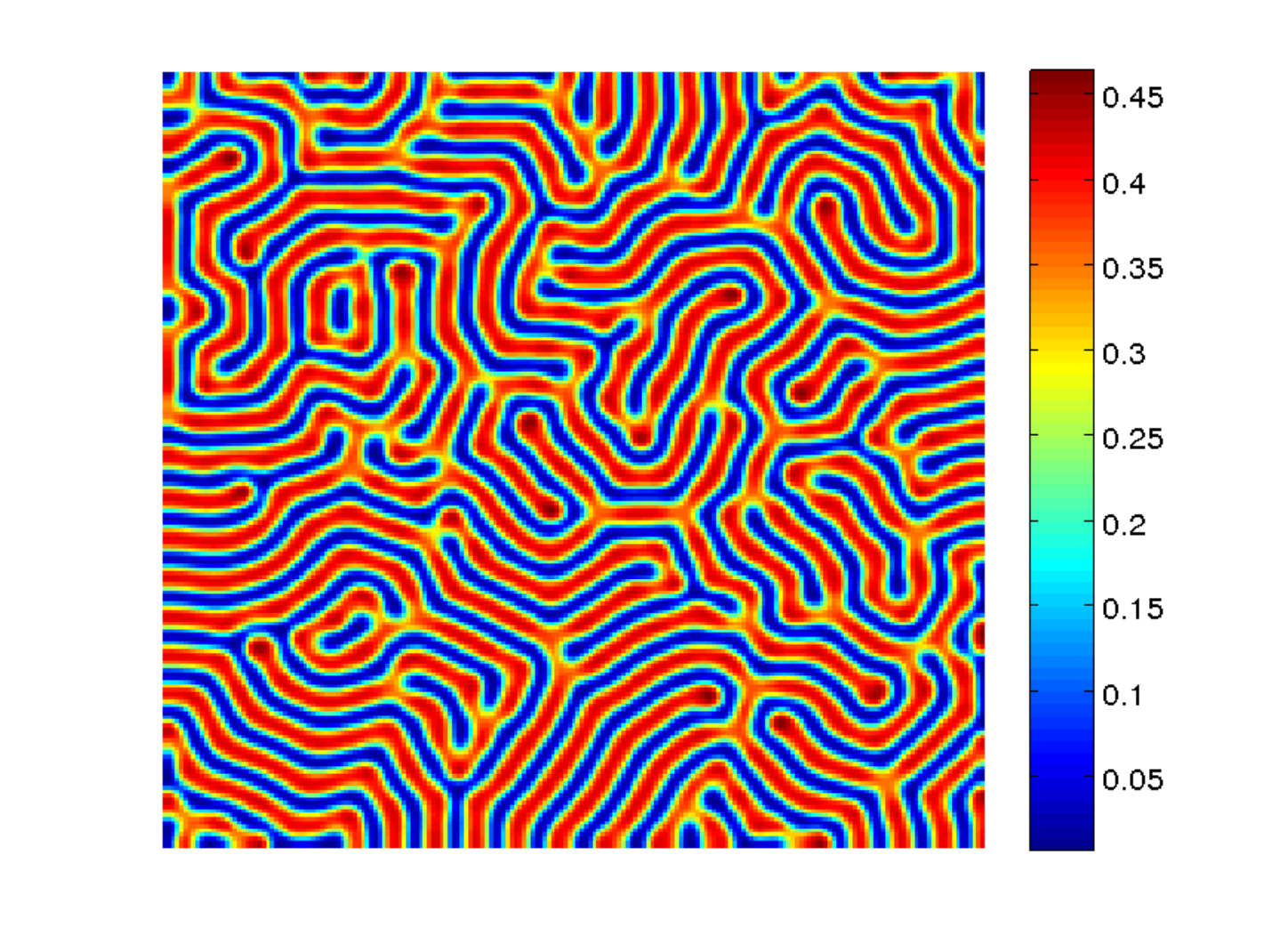}}
\subfigure[]{\includegraphics[width=2.5in,height=2in]{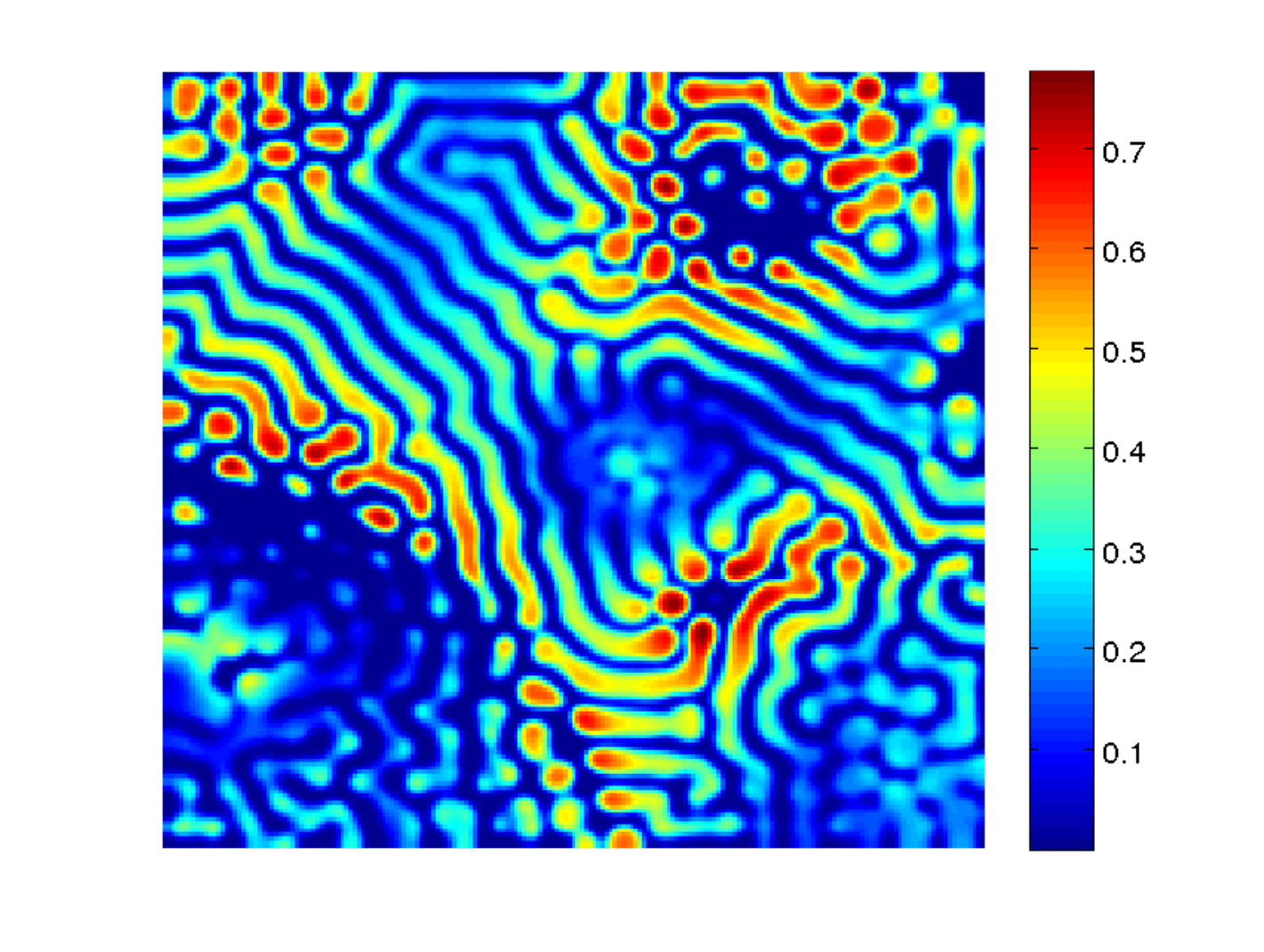}}}
\mbox{\subfigure[]{\includegraphics[width=2.5in,height=2in]{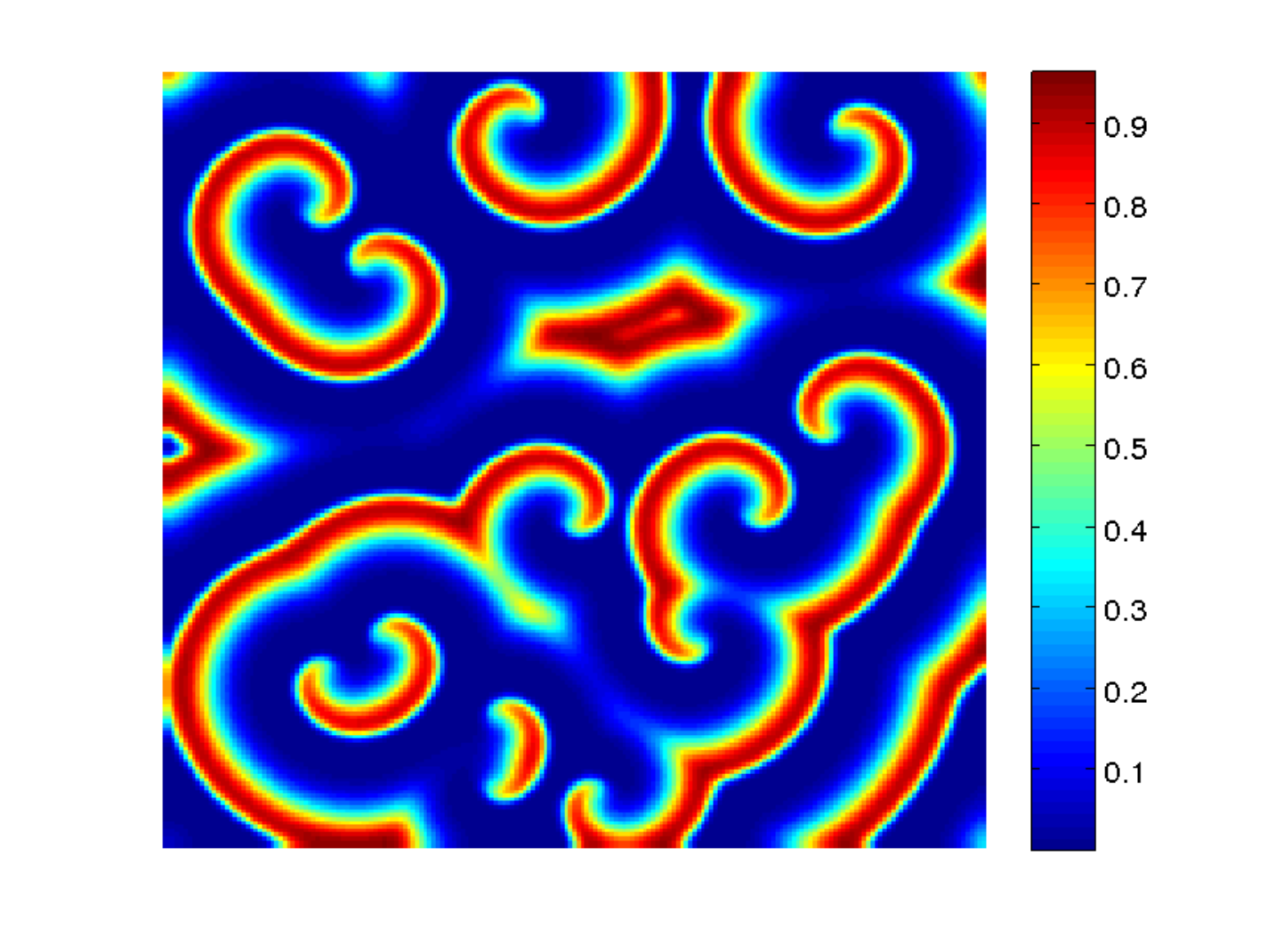}}}
\caption{\footnotesize Spatial distribution of prey ($ u $) obtained at
$t=1000$ for $\beta = 1$, $\gamma = 0.6$, $\delta = 0.1$, $d =10.0$
and (a) $ \alpha = 1.95$; (b) $ \alpha = 2.05 $; (c) $ \alpha = 2.15
$; (d) $ \alpha = 2.30$; and (e) $ d = 1.0$, $\alpha = 2.15$. The color bar in each panel represents the magnitude of $ u $.} \label{fig:2}
\end{figure}
%%%%%%%%%%%%%%%%%%%%%%%%%%%%%%%%%%%%%%%%%%%%%%%%%%%%%%%%%%%%%%%%

\par
\begin{figure}%
\leavevmode \centering
\mbox{\subfigure[]{\includegraphics[width=2.5in,height=2in]{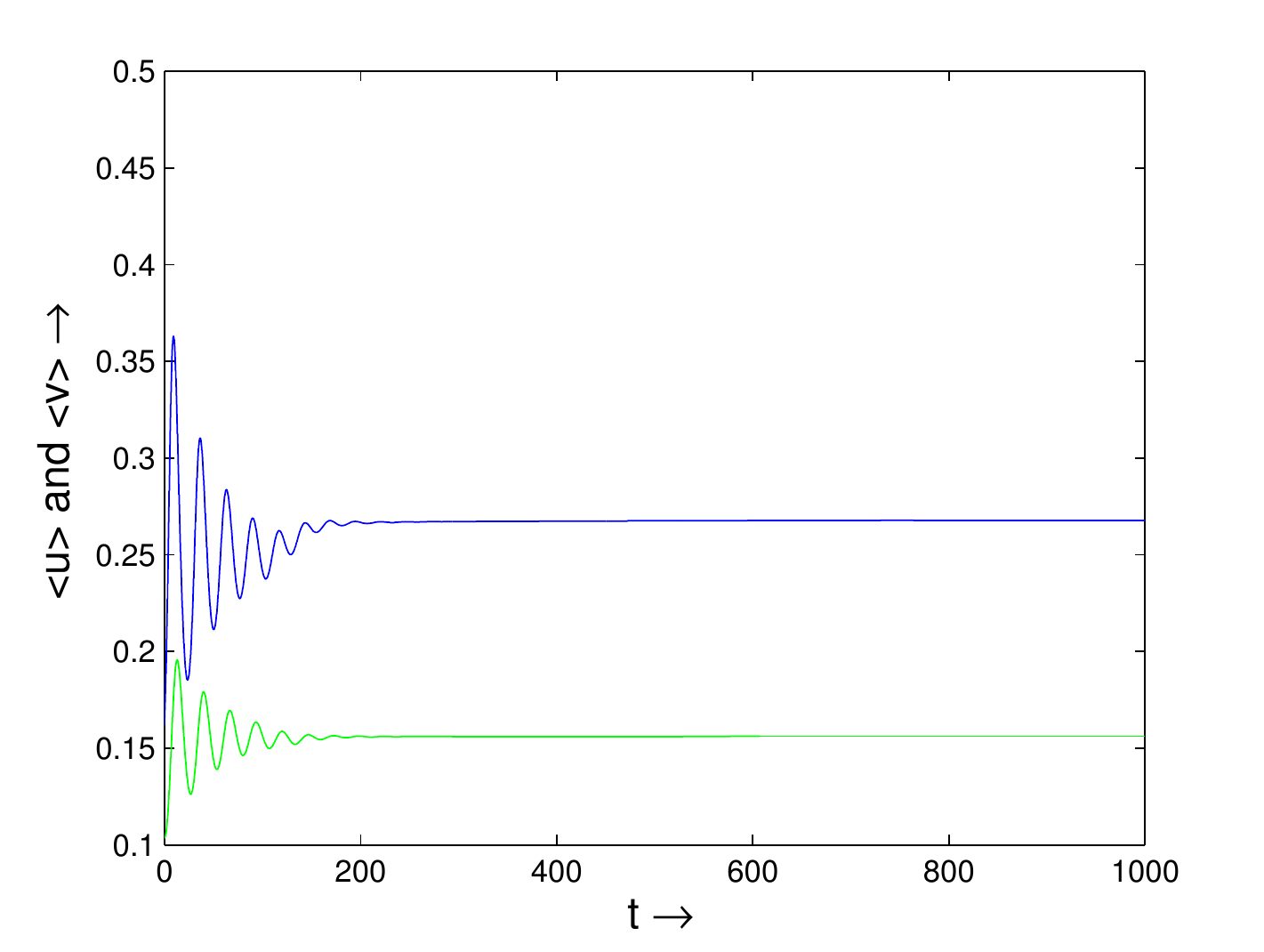}}
\subfigure[]{\includegraphics[width=2.5in,height=2in]{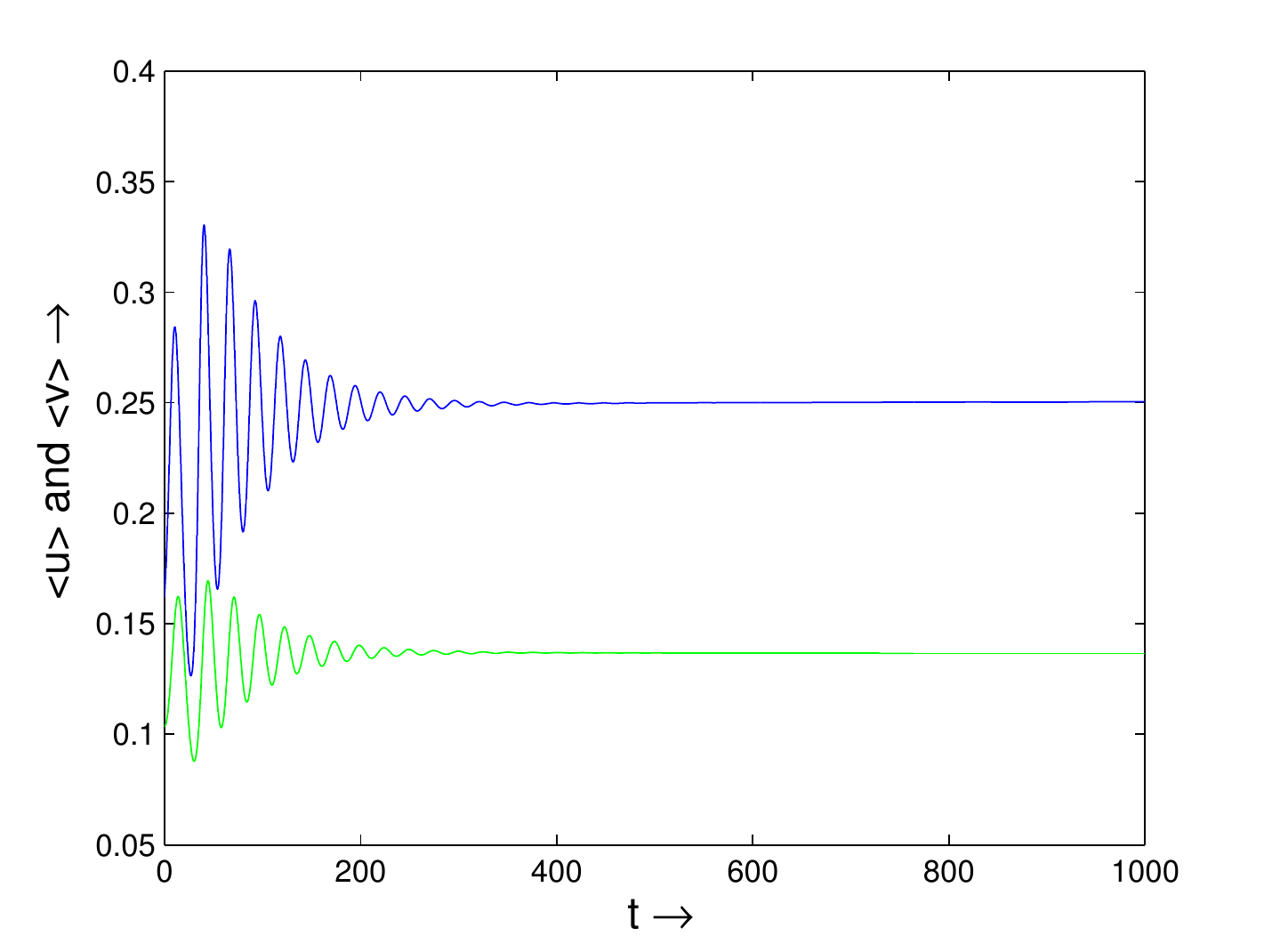}}}
\mbox{\subfigure[]{\includegraphics[width=2.5in,height=2in]{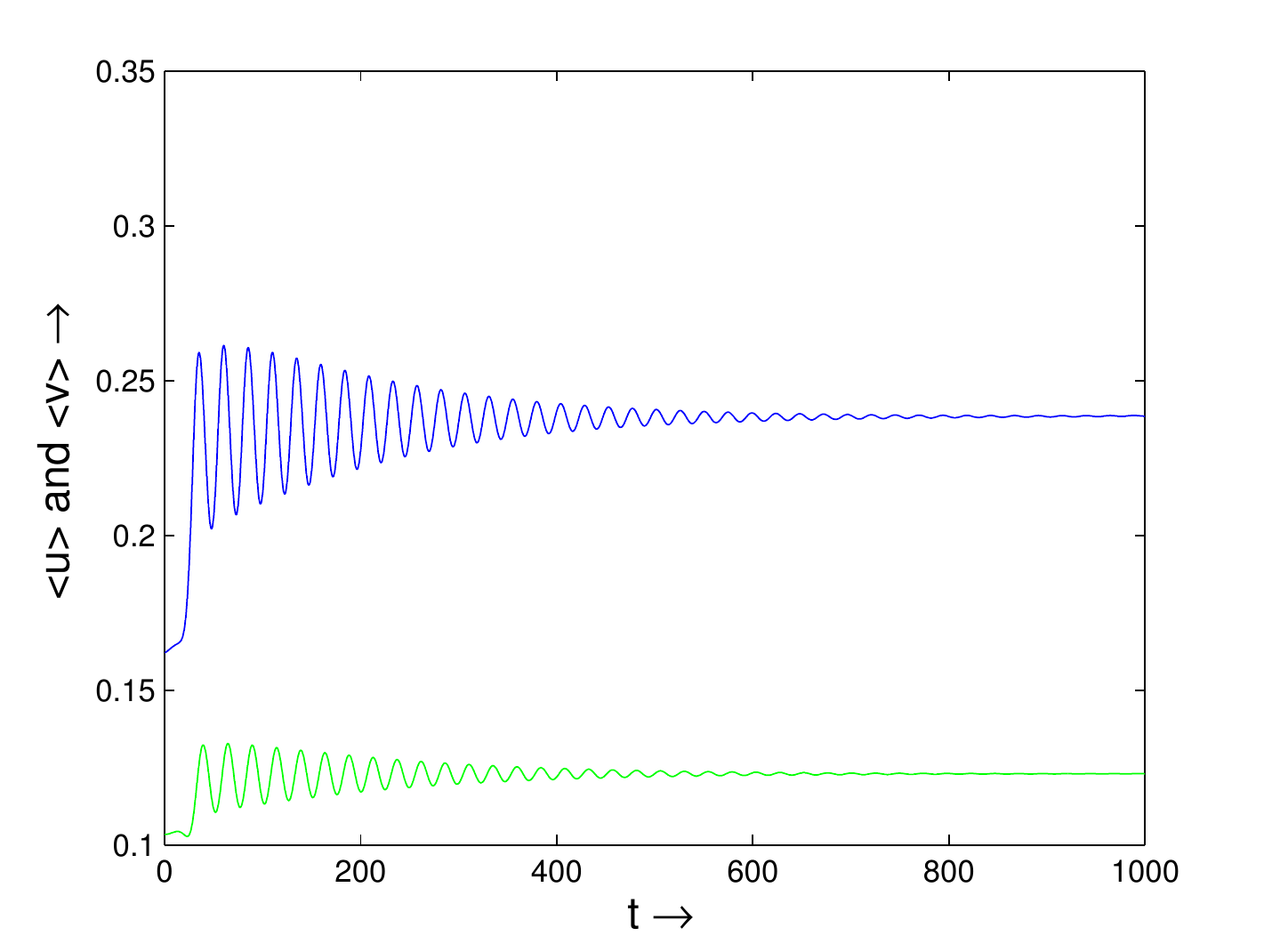}}
\subfigure[]{\includegraphics[width=2.5in,height=2in]{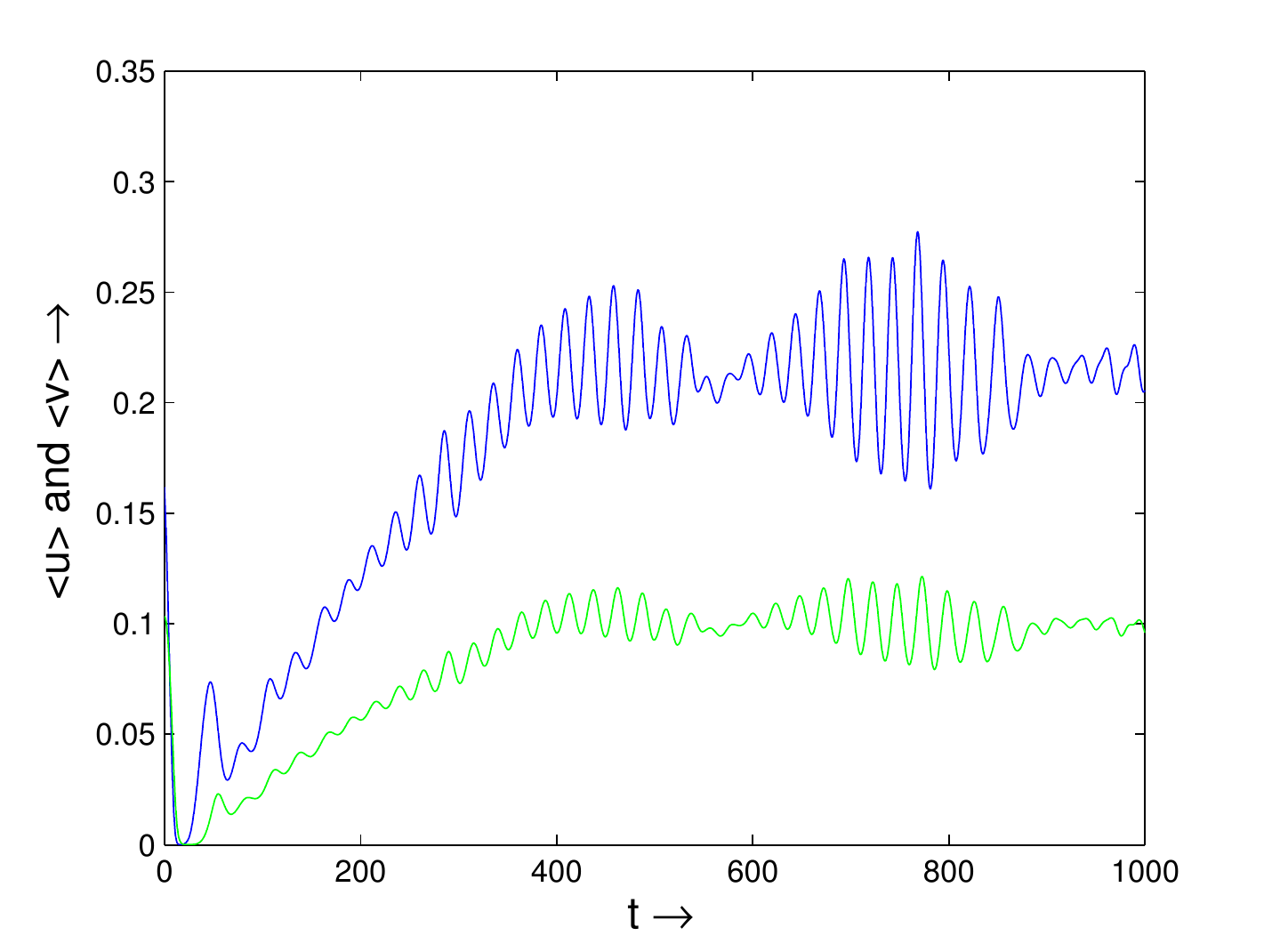}}}
\mbox{\subfigure[]{\includegraphics[width=2.5in,height=2in]{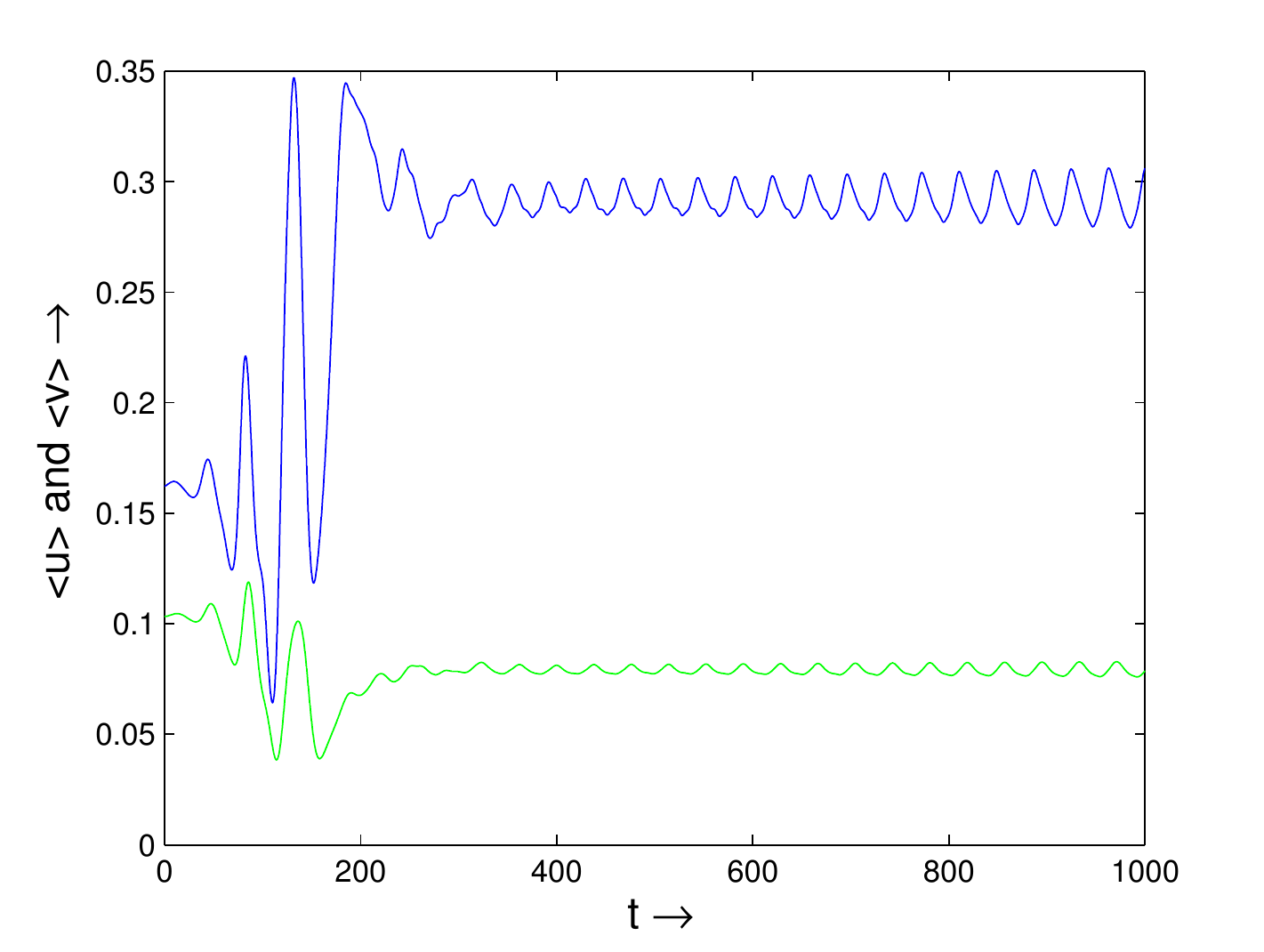}}}
\caption{\footnotesize Time evolution of spatial averages of the
prey $\left\langle u \right\rangle $ and predator $\left\langle v \right\rangle $ populations, presented in (dark) blue and and (light) green colors,
respectively, for $t\in[0,1000]$ corresponding to the patterns
presented in Fig.~\ref{fig:2}.}\label{fig:3}
\end{figure}

 A sample of each type of patterns observed within the Turing domain
are shown in Fig.~\ref{fig:2}. At the lower panel we have presented
the interacting spiral pattern which is observed for parameter
values just outside the Turing-Hopf domain and for $d=1.0$. As the
Turing instability can not occur for $d=1.0$, we can label this
pattern as non-Turing pattern. To understand the stationary nature
of the observed patterns, we have calculated the time evolution of
spatial average for the prey and predator densities against time
corresponding to the patterns reported in Fig.~\ref{fig:2} and the
results are presented in Fig.~\ref{fig:3}. The time evolution of
$\left\langle u \right\rangle $ and $\left\langle v \right\rangle $ are considered over a longer period of time to
ensure that the final observations are far away from the transient
behavior. Looking at the time evolution of spatial averages
presented at Fig.~\ref{fig:3}, one can easily understand that
cold-spot, mixture of spot-stripe and labyrinthine patterns are
stationary patterns whereas the interacting spiral pattern is a
non-stationary pattern. The time evolution of the spatial averages
are continuously oscillating and never settle down to any stationary
value. Corresponding to the spatiotemporal chaotic pattern presented
in Fig.~\ref{fig:2}(d), the time evolution of spatial averages
exhibit irregular oscillations for $t>650$ and the irregularity
sustains at all future time. Here we have presented the results up
to $t=1000$ from the sake of brevity.

%%%%%%%%%%%%%%%%%%%%%%%%%%%%%%%%%%%%%%%%%%%%%%%%%%%%%%%%%%%%%%%%
%%%%%%%%%%%%%%%%%%%%%%%%%%%%%%%%%%%%%%%%%%%%%%%%%%%%%%%%%%%%%%%%

\section{Noise Added Model}
\par In this section we are interested to study the effect of noise
on the spatiotemporal pattern formation for the model
(\ref{eq1}-\ref{eq2}). For this purpose we introduce uncorrelated
multiplicative white noise terms to the growth equations of both the
prey and the predator populations, accordingly the noise added model
is given by:
\begin{eqnarray}\label{noisemodel1}
\frac{\partial u}{\partial t}=& u\left(1-u\right) - \frac{\alpha uv}{u+v} + \nabla^2 u + \sigma_{1}u\xi_{1}(t,x,y),\\
\label{noisemodel2} \frac{\partial v}{\partial t}=& \frac{\beta
uv}{u+v} - \gamma v -\delta v^2 + \nabla^2 v +
\sigma_{2}u\xi_{2}(t,x,y),
\end{eqnarray}
where $\xi_{1}(t,x,y)$ and $\xi_{2}(t,x,y)$ are two temporally as
well as spatially uncorrelated Gaussian white noise terms. The
averages and correlations of the noise functions are given by,
$$ \left\langle \xi_{1}\left(t,x,y\right)\right\rangle = \left\langle\xi_{2}\left(t,x,y\right)\right\rangle = 0$$
$$\left\langle \xi_{j}\left(t,x,y\right)\xi_{j}\left(t_{1},x_{1},y_{1}\right)\right\rangle
= 2\sigma_{j}\delta\left(t - t_{1}\right)\delta\left(x -
x_{1}\right)\delta\left(y - y_{1}\right), j = 1,2,$$ where $
\sigma_{1}$ and $\sigma_{2}$ are the intensities of the
environmental driving forces. The introduction of the noise terms
can be justified from ecological point of view also. Here we are
considering the dimensionless model and hence we can see from
(\ref{eq1}-\ref{eq2}) that the intrinsic growth rate of the prey is
`1' and the intrinsic death rate for the predators is `$\gamma$'.
Now we assume that the intrinsic growth rate of the prey and the
death rate of the predator population are subjected to environmental
fluctuations and hence we can perturb them by spatiotemporal noise
terms. As a result, if we consider the introduction of noise terms
into the said rate constants, that is, $1\rightarrow
1+\sigma_1\xi_1(t,x,y)$ and $\gamma\rightarrow
\gamma+\sigma_2\xi_2(t,x,y)$ then we find the noise added model
(\ref{noisemodel1}) - (\ref{noisemodel2}) from the original
spatiotemporal model (\ref{eq1}-\ref{eq2}).

%%%%%%%%%%%%%%%%%%%%%%%%%%%%%%%%%%%%%%%%%%%%%%%%%%%%%%%%%%%%%%%%%
\begin{figure}
\leavevmode \centering
\mbox{\subfigure[]{\includegraphics[width=7.0cm]{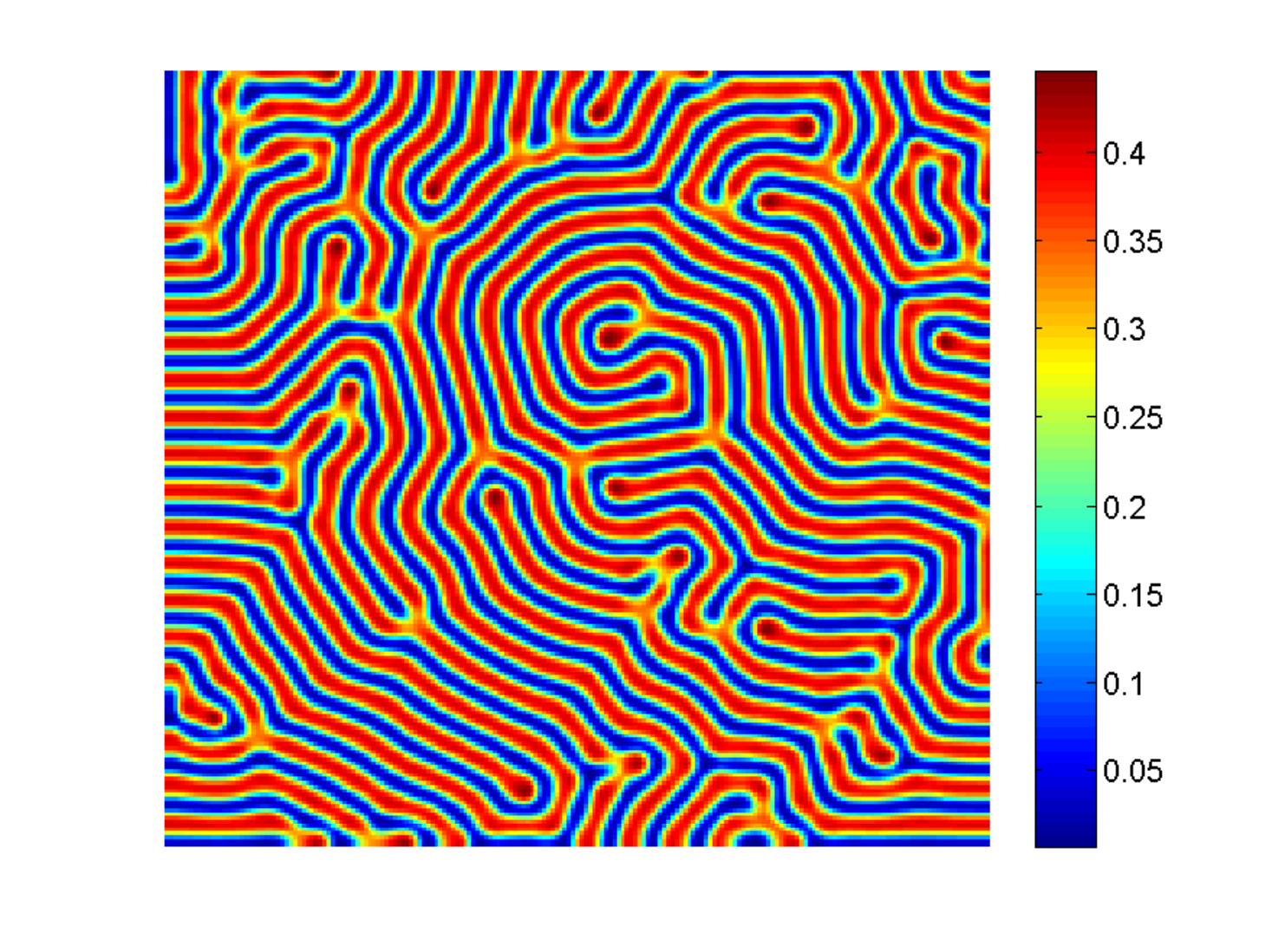}}
\subfigure[]{\includegraphics[width=7.0cm]{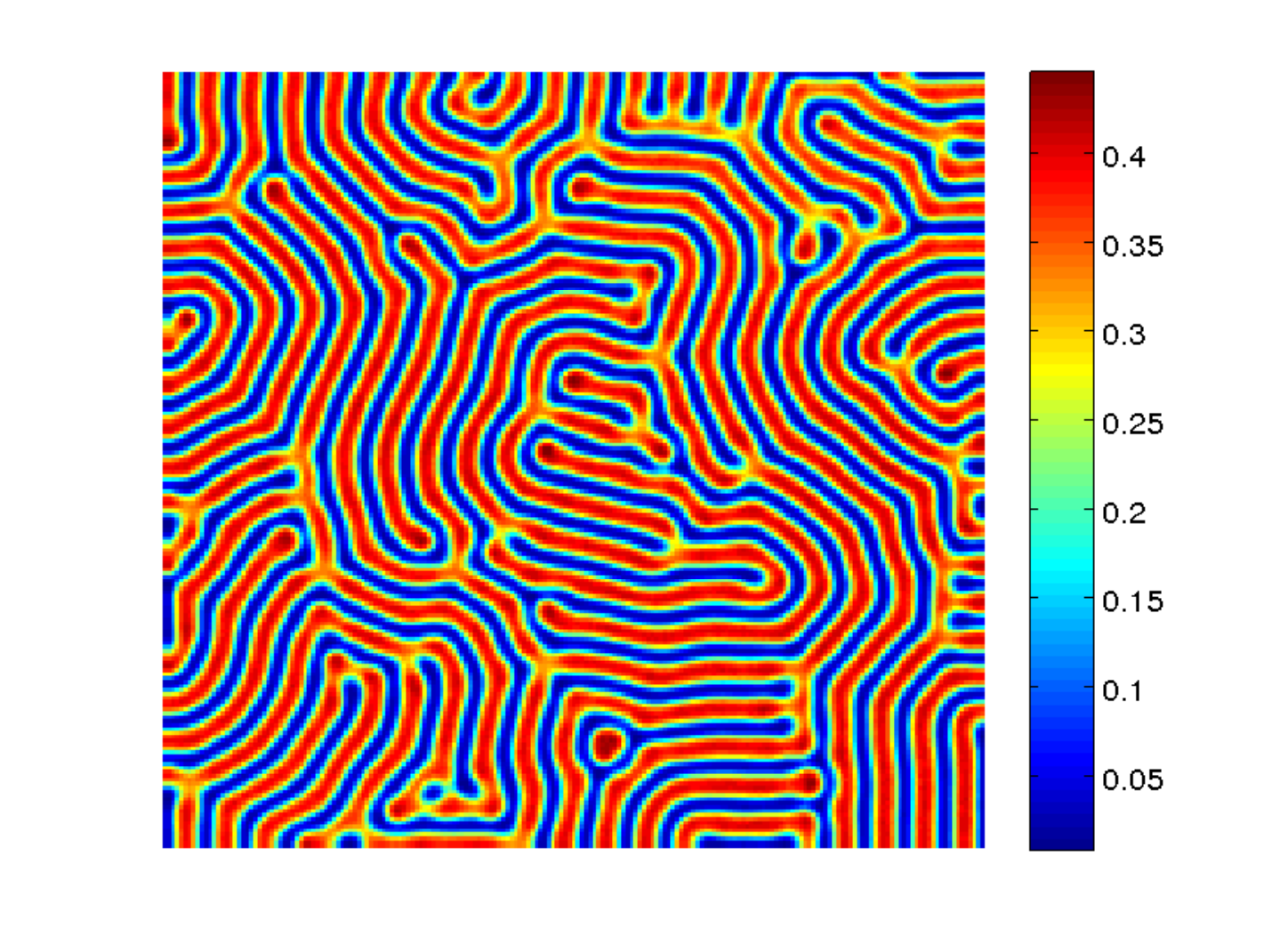}}}
\mbox{\subfigure[]{\includegraphics[width=7.0cm]{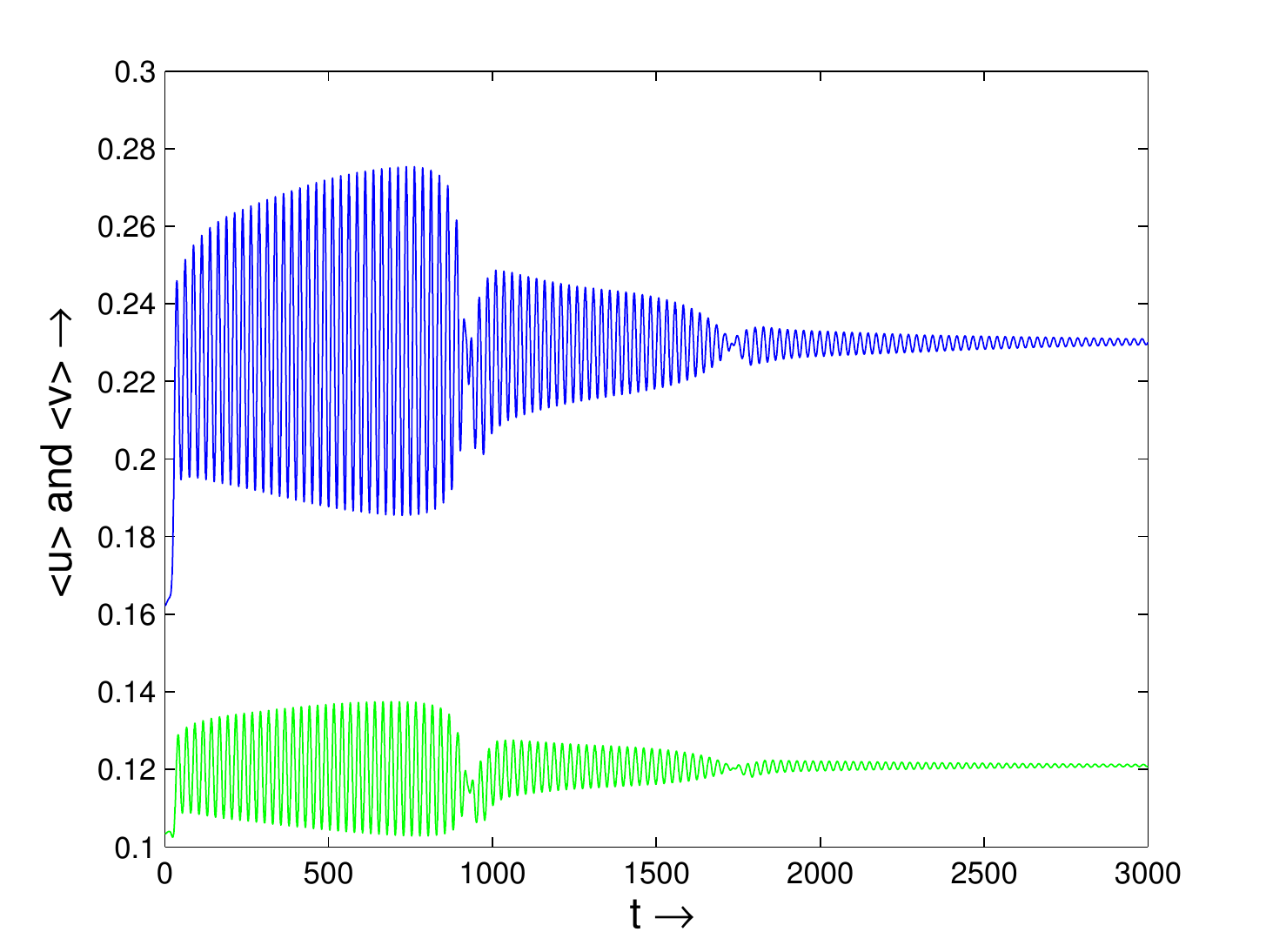}}
\subfigure[]{\includegraphics[width=7.0cm]{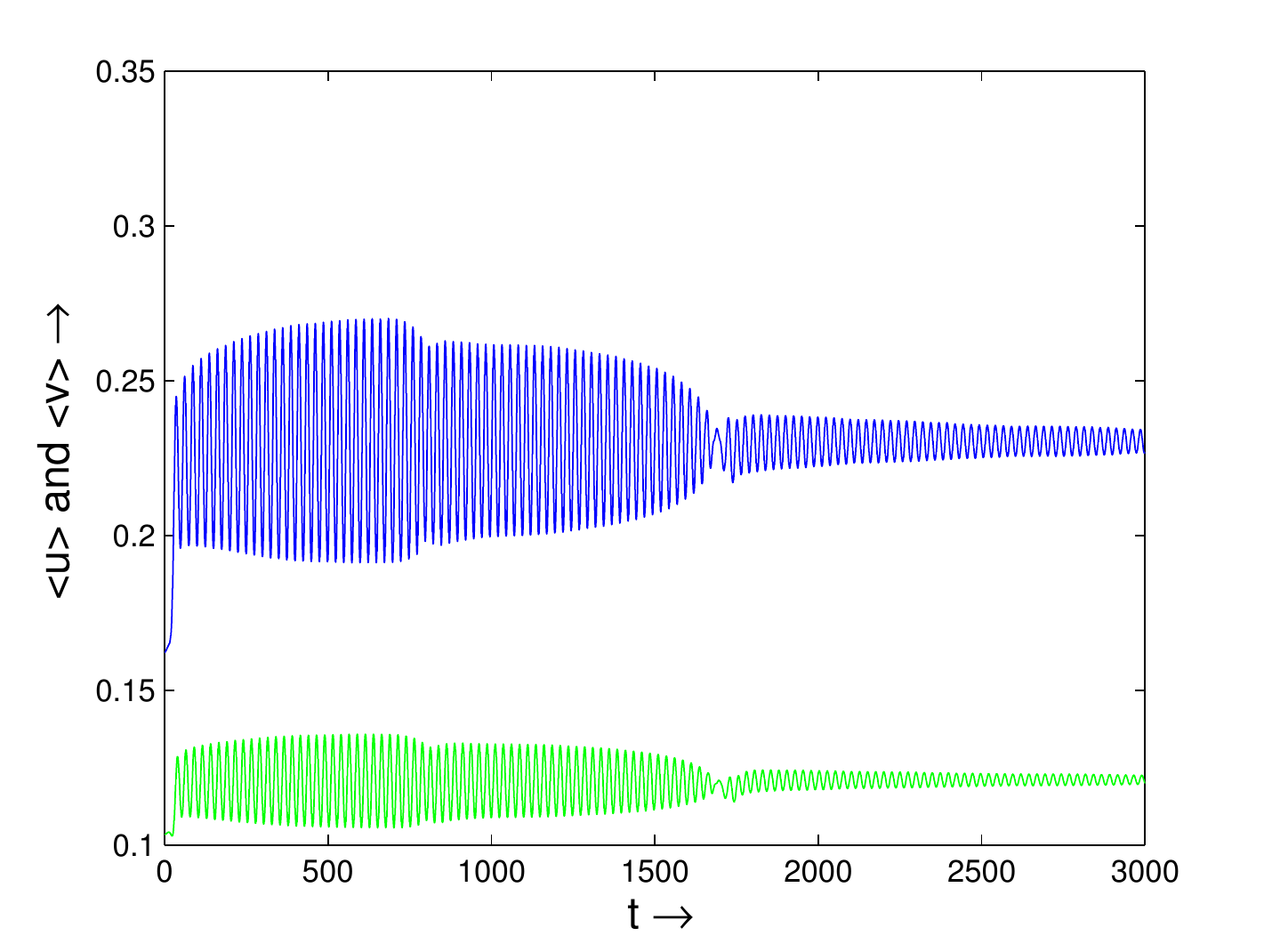}}}
\caption{\footnotesize Spatial distribution of prey obtained at
$t=3000$ for $\alpha = 2.15, \beta = 1,\gamma = 0.6, \delta = 0.1$
and $ d =8.5$ (a) $\sigma = 0$ (b) $\sigma = 0.01$; the color bar in each panel represents the magnitude of $ u $. Time evolution
of spatial averages of prey $\left\langle u \right\rangle $ and predator $\left\langle v \right\rangle $ populations   in (dark) blue and and (light) green colors,
respectively, for (c) $\sigma = 0$, and (d)
$\sigma = 0.01$.} \label{fig:4}
\end{figure}
%%%%%%%%%%%%%%%%%%%%%%%%%%%%%%%%%%%%%%%%%%%%%%%%%%%%%%%

\par In order to understand the effect of noise on the spatiotemporal
pattern formation, we have performed numerical simulations with the
changing magnitude of noise intensities. Here we consider the
parameter set $\alpha = 2.15$, $\beta = 1$, $\gamma = 0.6$, $\delta
= 0.1$, $ d = 8.5$ and assume that both the noise intensities are
same, that is $\sigma_{1}= \sigma_{2} = \sigma$. The numerical
simulation is performed for $\sigma=0.01$ and the resulting
distribution of the prey population is shown at Fig.~\ref{fig:4}(b)
along with the pattern observed for $\sigma=0$ (see
Fig.~\ref{fig:4}(a)). There is no significant difference in the
reported patterns as the noise intensity is small. We have also
calculated the spatial averages for the prey and predator
populations up to $t=3000$ for $\sigma=0$ and $\sigma=0.0010$ and
the results are presented at Fig.~\ref{fig:4}(c)-(d). It is clear
from these figures that the small noise does not change the system
characteristics when the spatiotemporal patterns are oscillatory in
nature. One can observe similar results for the parameter values
corresponding to the stationary patterns and when noise intensities
are not very high.

\par
\begin{figure}
\leavevmode \centering
\mbox{\subfigure[]{\includegraphics[width=7.0cm]{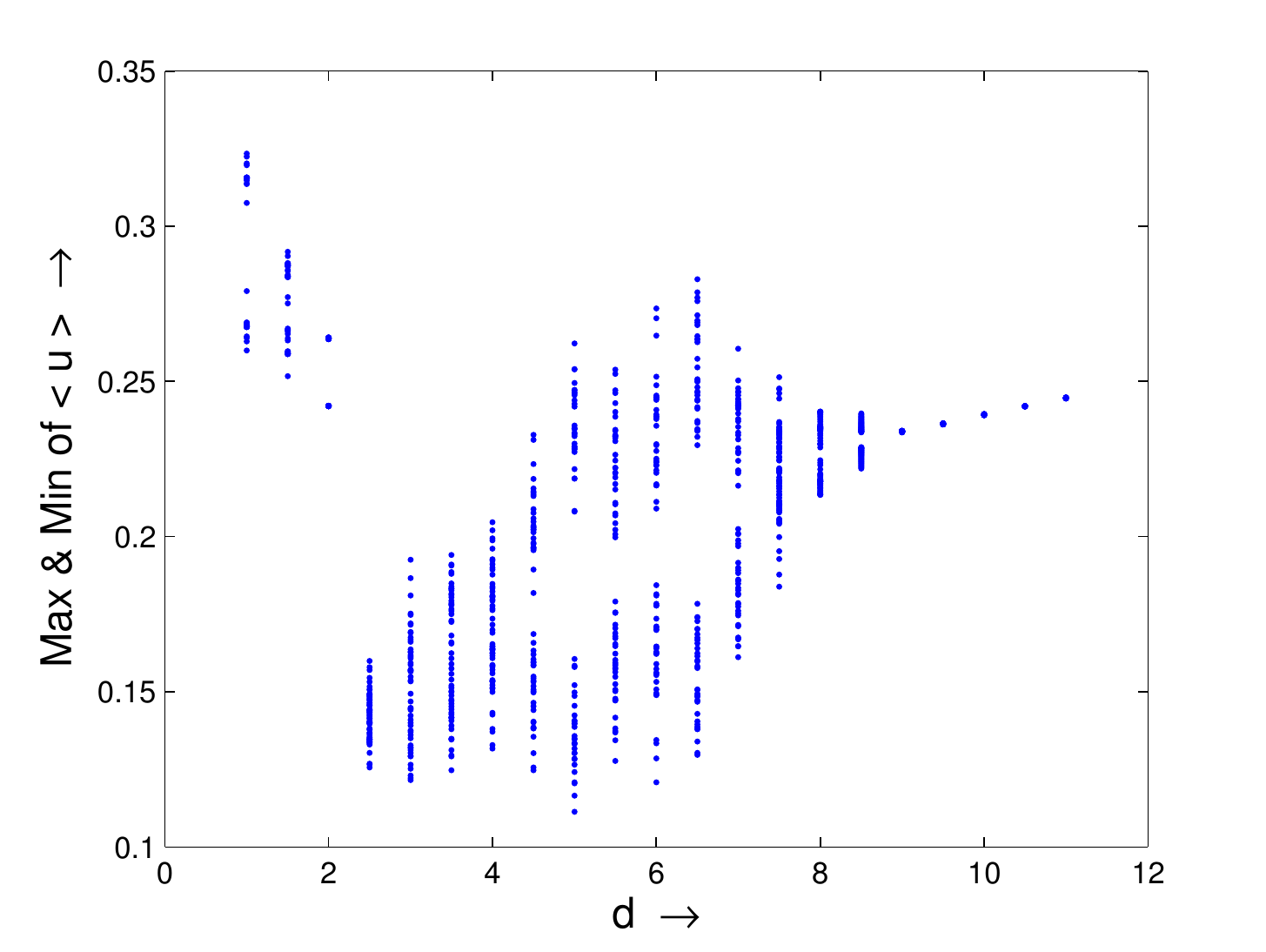}}
\subfigure[]{\includegraphics[width=7.0cm]{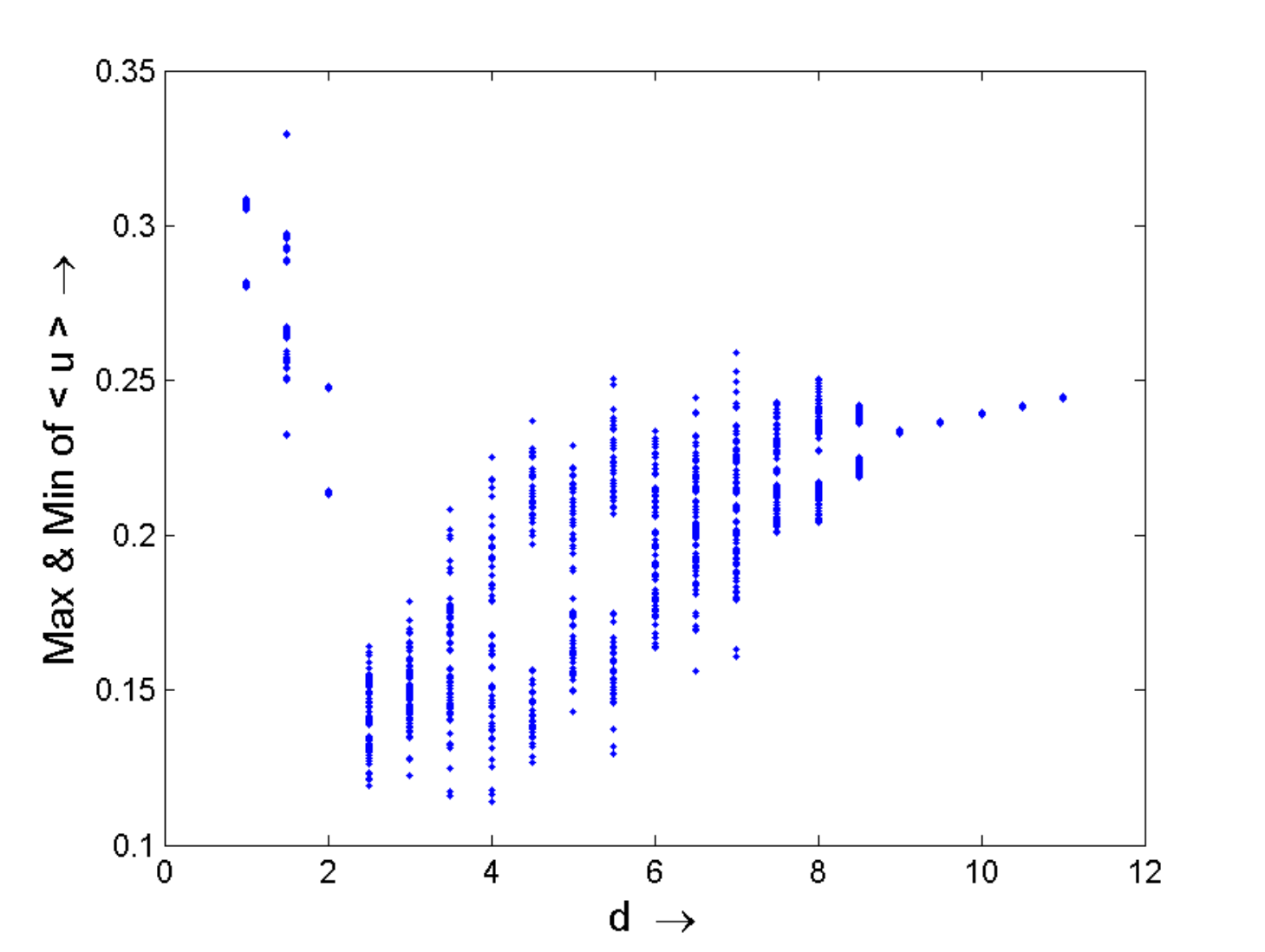}}}
\caption{\footnotesize Bifurcation diagram with respect to
bifurcation parameter $d$ with other parameters remaining same as in the
previous figure for (a) $\sigma = 0$ and (b) $\sigma = 0.01$.}
\label{fig:5}
\end{figure}

 In order to support our claim, we have prepared a bifurcation
diagram for the same set of parameter values, as mentioned above,
and considering the ratio of diffusivity as the bifurcation
parameter. The bifurcation diagrams are prepared for the
spatiotemporal model (\ref{eq1}-\ref{eq2}) as well as for the noise
added model (\ref{noisemodel1}) - (\ref{noisemodel2}) with
$\sigma=0.01$, see Fig.~\ref{fig:5}. We have performed the
numerical simulations for $d=1.0$ to $d=11.0$ with increments of $0.5$ and
spatial averages are calculated for the prey population up to
$t=3000$. After discarding the initial transients, all local maxima
and minima of the spatial average of the prey population are plotted
against each values of `$d$'. If we compare the bifurcation
diagrams, then it is evident that the number of local maxima and
minima increases with the introduction of small amplitude noise
terms mainly within the parameter regime where we find
non-stationary spatial patterns in the absence of noise. One natural
question arises whether the bifurcation diagram presented at the
right panel of Fig.~\ref{fig:5} changes from one simulation to
other. It does not change as we have taken the spatial average at
each time step and the spatial domain is sufficiently large to avoid
such kind of variation in the bifurcation diagram.

%%%%%%%%%%%%%%%%%%%%%%%%%%%%%%%%%%%%%%%%%%%%%
%%%%%%%%%%%%%%%%%%%%%%%%%%%%%%%%%%%%%%%%%%%%%%%%%%%%%%%%%%%%%%%%%

\par
\begin{figure}
\leavevmode \centering
\includegraphics[width=7.0cm]{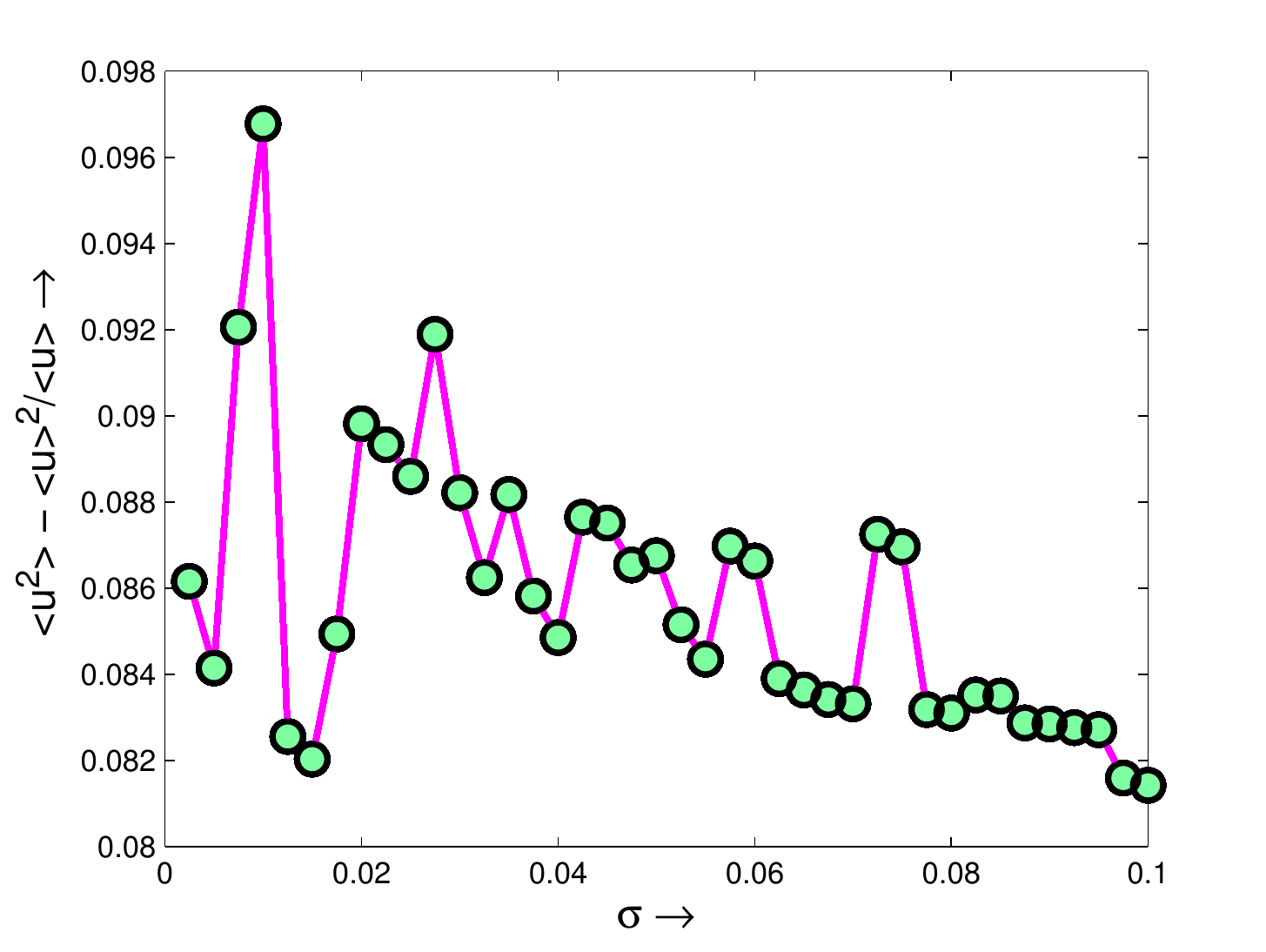}
\caption{\footnotesize Signal to Noise ratio measure for $ \alpha =
2.15$, $\beta = 1$, $\gamma = 0.6$, $\delta = 0.1$, $d = 8.5$. } \label{fig:6}
\end{figure}

 After understanding the effect of small intensity noise on the
resulting patterns, it is important to understand what happens with
the increasing magnitude of noise intensities. Of course, we can
find some noisy patterns when the magnitude of $\sigma$ is
significantly large and some times it is difficult to identify the
particular type of patterns they are representing. In order to avoid
this type of confusion and appropriate understanding of the
underlying phenomena, we have calculated the signal-to-noise-ratio
(SNR) against a range of noise intensities. Keeping $ \alpha =
2.15$, $\beta = 1$, $\gamma = 0.6$, $\delta = 0.1$, $d = 8.5$ fixed,
we have performed the numerical simulations for a range of values of
$\sigma$. After discarding the initial transients, we have
calculated the signal-to-noise-ratio measure $ \frac{<u^{2}> -<u>^{2}}{<u>}$ for $u(t,x,y)$ obtained
at $t = 3000$ and the average is taken over the entire spatial
domain \cite{pikovsky1997coherence}. The plot of the SNR measure is shown at
Fig.~\ref{fig:6} when $\sigma$ varies within the range
$[0.0025,0.1]$. From the figure, it is evident that the relative
variance of the population is not a monotonic increasing function of
the noise intensities. The environmental driving force can cause the
dispersion of the populations from their average value but the
internal mechanism can compensate to maintain the size of the
population patches and as a result we find that the relative
variance is an increasing function of the noise intensity.

%%%%%%%%%%%%%%%%%%%%%%%%%%%%%%%%%%%%%%%%%%%%%%%%%%%%%%%%%%%%%%%%%
%%%%%%%%%%%%%%%%%%%%%%%%%%%%%%%%%%%%%%%%%%%%%%%%%%%%%%%

\section{Conclusion}

In this paper, we have studied a predator-prey model with ratio-dependent functional response and density dependent death rate of predator, incorporated with the self-diffusion terms corresponding to the random movement of the individuals within two dimensional habitats. Extensive numerical simulations were performed to understand the regimes of patterns that the model exhibited within and outside of the Turing boundary.  Typically, four types of patterns were observed within the Turing domain and interacting spiral pattern as non-Turing pattern. The effect of environmental driving forces (noise) were also studied. Interestingly, we  observed that the small amplitude noise  does not alter the stationary pattern; only the time taken to reach the stationary pattern increases. However, irregularity increases even with small noise intensities within the non-stationary regime. We observed that dispersion of population from their spatial average does not increase gradually with the noise intensities; rather noise and demographic interaction play crucial roles to determine the distribution of populations within their habitats, even during the adverse environmental conditions. Our next attempt will be to address the issues of parameter values within the spatiotemporal chaotic domain, and also investigate different noise scenarios such as  $\sigma_1\neq \sigma_2$.

%%%%%%%%%%%%%%%%%%%%%%%%%%%%%%%%%%%%%%%%%%%%
\section*{References}

\bibliographystyle{iopart-num}
\bibliography{jpcs_AKS_MB_AC}

\end{document}